\documentclass[aps,prl,twocolumn,showpacs,10pt,superscriptaddress,preprintnumbers,footinbib]{revtex4-1}
\usepackage{amsmath}
\usepackage{graphicx}
\usepackage{amssymb}
\usepackage[colorlinks,citecolor=red]{hyperref}
\usepackage{slashed}
\usepackage{color}
\usepackage{ulem}
\usepackage{multirow}
\usepackage{booktabs}
\usepackage{ulem}
\usepackage{slashed}

\newcommand{\beq}{\begin{equation}}
\newcommand{\eeq}{\end{equation}}
\newcommand{\bqa}{\begin{eqnarray}}
\newcommand{\eqa}{\end{eqnarray}}
\newcommand{\nnb}{\nonumber\\}
\begin{document}
\title{Probing Glueball content of X(2370) through heavy quarkonium radiative decays and electron-positron annihilation}
\author{Ruilin Zhu}
\email{rlzhu@njnu.edu.cn}
\affiliation{Department of Physics and Institute of Theoretical Physics, Nanjing Normal University, Nanjing,
Jiangsu 210023, China}
\affiliation{Nanjing Key Laboratory of Particle Physics and Astrophysics, Nanjing Normal University, Nanjing,
Jiangsu 210023, China}
\author{Qiang Zhao}
\email{zhaoq@ihep.ac.cn}
\affiliation{Institute of High Energy Physics, Chinese Academy of Sciences, Beijing 100049, China}
\affiliation{University of Chinese Academy of Sciences, Beijing 100049, China}

\begin{abstract}
In this paper, we first analyze the light-cone distribution amplitudes of the pseudoscalar Glueball and point out that the two-gluon and quark-antiquark pair contributions naturally mix at the quantum loop level. Based on the $\eta-\eta'-G-\eta_c$ tetra-mixing scheme  and existing experimental results, we analyze the branching ratio of $J/\psi \to P+\gamma$ with pseudoscalar meson $P$ and extract the mixing parameters of the tetra-mixing scheme. We then successfully predict the branching ratio of $\psi(2S) \to P+\gamma$, which is found to be in good agreement with experimental results. We also find that the branching ratios of $J/\psi \to X(2370)+\gamma$ and $\psi(2S) \to X(2370)+\gamma$  are highly sensitive to the Glueball-charmonium mixing angle: the destructive mixing interference in the branching ratio of $J/\psi \to X(2370)+\gamma$ occurs at
$\phi_{gc}\approx0.01$, whereas that in $\psi(2S) \to X(2370)+\gamma$  occurs at $\phi_{gc}\approx 0.11$. Moreover, the $J/\psi \to X(2370)+\gamma$ branching ratio exhibits constructive mixing interference when the mixing angle is larger than $\phi_{gc}\approx0.02$. Finally, we also study the electroproduction cross sections of pseudoscalar mesons. After excluding the resonance contributions, the theoretical calculations agree with the experimental measurements in order of magnitude, and we also present theoretical predictions for the electroproduction cross section of the $X(2370)$. These theoretical results can be readily tested by current BESIII and Belle-II experiments, and further experimental measurements will reveal the intrinsic nature of the $X(2370)$.
\end{abstract}

\maketitle

\textit{Introduction.}
The search for Glueball is a major problem in particle physics and its confirmation will deepen our understanding of non-Abelian gauge fields and the color confinement mechanism. Very recently, the lightest pseudoscalar Glueball has been proposed as the dominant constituent of the $X(2370)$, based on a systematic analysis by the BESIII Collaboration of its mass, spin-parity, and production rate~\cite{BESIII:2026mvn}. The $X(2370)$ was first discovered in the invariant mass spectrum of $\pi^+\pi^-\eta'$ of the $J/\psi$ radiative decay fifteen years ago~\cite{BESIII:2010gmv}, where its mass and width were measured to be $2376.3\pm8.1(stat)^{+3.1}_{-4.3}(syst) MeV$ and $83\pm17(stat)^{+44}_{-6}(syst) MeV$. Twelve years later, the spin-parity quantum numbers of the $X(2370)$ was determined as $J^{PC}=0^{-+}$ by the partial wave analysis~\cite{BESIII:2023wfi}. An important observation in recent experiment is that the suppression of the $ X(2370)\to K^*(892)\overline{K}$ mode while the order of a few times
$10^{-4}$ for decay rates of $J/\psi \to \gamma X(2370)\to \gamma K\overline{K}\pi,  \gamma \pi\pi\eta,  \gamma \pi\pi\eta'$ modes, which indicated that the $X(2370)$ is mostly like a flavor-singlet state~\cite{BESIII:2026mvn}.

Starting from the fundamental theory of Quantum Chromodynamics (QCD), the lightest pseudoscalar Glueball mass lies between 2.1GeV and 2.7GeV after taking the uncertainties into account from several lattice QCD simulations~\cite{Morningstar:1999rf,Chen:2005mg,Gregory:2012hu}, where
the $X(2370)$ mass is just within that range. From the mass, the spin-parity quantum numbers, the production and decay properties of the $X(2370)$,
this state is mostly like the lightest pseudoscalar Glueball~\cite{BESIII:2026mvn,Huang:2025pyv}. Recent theoretical explanation of the $X(2370)$ as a pure Glueball from QCD sum rules are in Refs.~\cite{Wang:2026iyq,Tan:2026crv}. However, the fundamental interactions between quarks and gluons in QCD will produce sea quarks or more gluons, which may mix between Glueball and quark-antiquark states with identical spin-parity quantum numbers.

Therefore, it is natural to ask how large the pseudoscalar Glueball component in the $X(2370)$ is? This question can be precisely determined from its production and decay properties of the $X(2370)$. From the latest experimental measurements ${\cal B}(J/\psi \to \gamma X(2370)\to \gamma K\overline{K}\pi)=(3.25\pm0.25^{+0.73}_{-0.75})\times 10^{-4}$, ${\cal B}(J/\psi \to \gamma X(2370)\to\gamma \pi\pi\eta)=(3.2\pm0.1^{+0.9}_{-1.0})\times 10^{-4}$,  ${\cal B}(J/\psi \to \gamma X(2370)\to\gamma \pi\pi\eta')=(1.94\pm0.04^{+0.33}_{-0.88})\times 10^{-4}$  and  ${\cal B}(J/\psi \to \gamma X(2370)\to\gamma K\overline{K}\eta')=(0.39\pm0.05\pm0.10)\times 10^{-4}$, a large decay rate ${\cal B}(J/\psi \to \gamma X(2370)$ is estimated as around $10^{-3}$. However, the radiative decay rate of $J/\psi$ into a pseudoscalar Glueball has been predicted as ${\cal B}(J/\psi \to \gamma G(J^{PC}=0^{-+})=(2.3\pm0.8)\times 10^{-4}$ from lattice QCD simulation~\cite{Gui:2019dtm}, where a pure pseudoscalar Glueball explanation of the $X(2370)$ alone cannot explain the current experimental data. Very recently, the small angle mixing scheme between the Glueball-charmonium  can produce  a large production rate in the radiative decay of $J/\psi$ into a pseudoscalar Glueball~\cite{Chen:2026lki}.

In this paper, we study the $X(2370)$ production in heavy quarkonium radiative decays and electron-positron annihilation processes from QCD factorization theory. First we give the definition of the light-cone distribution amplitude (LCDA) for Pseudoscalar Glueball. Then we revise
the $\eta-\eta'-G-\eta_c$ tetra-mixing scheme and treat the $X(2370)$ as one of the tetra-mixing products. We calculate the $J/\psi$ radiative decay to pseudoscalar meson  and extract the independent mixing angles based on the experimental data. In the end, we predict the decay rates for $\psi(2S)$, $\Upsilon(1S)$ and $\Upsilon(2S)$ radiative decay to pseudoscalar meson and the production cross section for pseudoscalar meson in electron-positron annihilation process.

\begin{figure}[th]
\begin{center}
\includegraphics[width=0.4\textwidth]{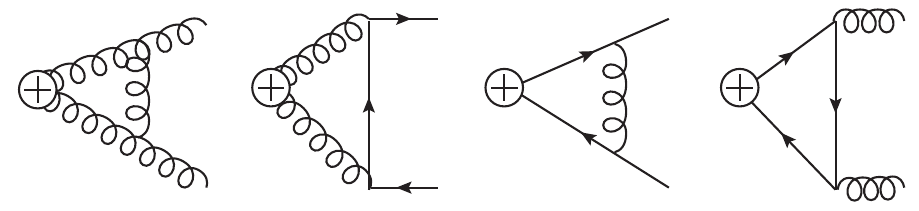}\,\,\,
\caption{Typical Feynman diagrams for the gluonium and quark-antiquark operators mixing, which start to contribute at loop level.}
\label{fig:feynmandiagmix}
\end{center}
\end{figure}

\textit{Pseudoscalar Glueball LCDA.}
Pseudoscalar Glueball is made of two constituent gluons with spin-parities quantum numbers $J^{PC}=0^{-+}$.
Due to the interactions between quarks and gluons in QCD, it is natural for sea quarks or more gluons to be produced and then mixed with Glueballs~\cite{Zhu:2015qoa,Zhu:2016arf}.
This mixing effect between the gluonium and flavor-singlet quark-antiquark operators with identical quantum numbers starts at Quantum loop level from Fig.~\ref{fig:feynmandiagmix}.

Thus it is easily to define a two dimensional light-cone distribution amplitude for Pseudoscalar Glueball
\beq
\mbox{\boldmath$\Phi$}(u)\equiv \left(
\begin{array}[c]{c}
\phi_{q}(u) \\[0.2em]
\phi_{g}(u)
\end{array}
\right) \, ,\label{eq:definition-LCDA}
\eeq
where the leading-order twist-2 LCDAs of $\phi_{i}(u)$ are
\begin{widetext}
\begin{eqnarray}
 \phi_q (u) &=& \int \frac{dz^-}{2\pi}N_q\frac{ e^{i(2u-1)k^+z^-/2 } }{ }\langle G(k)|\overline{\Psi}_i(-z^-/2)(\gamma_5\slashed{k})_{ik}L_{kj}(-z^-/2,z^-/2)\Psi_j(z^-/2)|0\rangle,\\
 \phi_g (u) &=& \int \frac{dz^-}{2\pi}
N_g\frac{ e^{i(2u-1)k^+z^-/2 } }{ u(1-u)}\epsilon_\perp^{\mu\nu}\langle G(k)|G^{a,+\mu}(-z^-/2)L_{ab}
(-z^-/2,z^-/2)G^{b,+\nu}(z^-/2)|0\rangle,
 \label{eq:definition-gauge-invariant-LCDA-qg}
\end{eqnarray}
\end{widetext}
where $N_{q,g}$ is the renormalization factor. The gauge link $L(x,y) = P\,e^{ig\int_0^1 ds (x-y)_\mu A^\mu((x-y)s+y)}$ in the fundamental and adjoint representation is to ensure the gauge invariant. The light-cone frame is employed by $k^\mu=(k^+,k^-,\mathbf{k}_\perp)$
with $k^+=(k^0+k^3)/\sqrt{2}$ and $k^-=(k^0-k^3)/\sqrt{2}$.
The projector $\epsilon_\perp^{\mu\nu}=\epsilon^{\alpha\beta\mu\nu}k_\alpha\bar{k}_\beta$ with $\bar{k}^\mu=(k^-,k^+,\mathbf{k}_\perp)$.
Due to the symmetry in LCDAs for pseudoscalar Glueball, we have
$\Phi_g(u,\mu^2)=-\Phi_g(1-u,\mu^2)$ and $\Phi_q(u,\mu^2)=\Phi_q(1-u,\mu^2)$.

\textit{The $\eta-\eta'-G-\eta_c$ mixing scheme.}
Due to the Chiral symmetry breaking, a large mixing angle between the SU(3) flavor-singlet and flavor-octet states are observed in previous studies~\cite{Ali:1997nh,Leutwyler:1997yr,Venugopal:1998fq,Feldmann:1998vh,Ball:2007hb,Agaev:2014wna,Alte:2015dpo,Gan:2025sac},
which is also employed to explain the $\eta'$ large mass. Meanwhile, the SU(3) flavor singlet-octet mixing scheme and $q\bar{q}-s\bar{s}$ mixing scheme are equivalent. In addition, $c\bar{c}$ is also a SU(3) flavor singlet and is natural to mix with gluonium, which is also observed at Lattice QCD simulation. Since $b\bar{b}$ is too heavy compared with Glueball, its mixing effect can be ignored. However, future precision experiments may be able to test the penta-mixing scheme that includes the $b\bar{b}$ state.

Following the $\eta-\eta'-G-\eta_c$ tetra-mixing scheme proposed by one of the authors~\cite{Tsai:2011dp,Peng:2011ue,Qin:2017qes}
\begin{align}
\left(\eta,\eta',G,\eta_c\right)^T=U_{4\times 4}(\Phi_i,\Gamma_j)\left(|q\bar{q}\rangle,|s\bar{s}\rangle,|gg\rangle,|c\bar{c}\rangle\right)^T,
\end{align}
where $i=1-6,j=1-3$ and there are 6 mixing angles and 3 phase angles in general.
However, these parameter values are hierarchical, and some of them can be completely neglected, as can be seen in previous studies of two-dimensional or three-dimensional mixing.

Combing previous studies and ignoring the phase angles, the $\eta-\eta'-G-\eta_c$ mixing matrix can be revised as
\begin{widetext}
\begin{align}
U_{4\times 4}\left(\phi,\theta_{g,c},a_{g,c},\phi_{gc}\right)=
\begin{pmatrix}
\cos \phi & -\sin \phi & -a_g\sin \theta_g &-a_c\sin \theta_c\\
\sin \phi & \cos \phi & a_g\cos \theta_g  &a_c\cos \theta_c\\
-a_g\sin \left(\phi-\theta_g\right) &-a_g\cos \left(\phi-\theta_g\right) & \cos\phi_{gc}&-\sin\phi_{gc}\\
-a_c\sin \left(\phi-\theta_c\right) &-a_c\cos \left(\phi-\theta_c\right) & \sin\phi_{gc}&\cos\phi_{gc}
\end{pmatrix}+{\cal O}\left(a_{g,c}^2,\theta_{g,c}^2,\phi_{gc}\right).
\end{align}
\end{widetext}

There are several possibilities to extract the values of these mixing angles from experiments.
The processes that can be referred to include $J/\psi \to \eta^{(')}+\gamma$, $J/\psi \to \eta^{(')}+\rho$, $\eta' \to \rho+\gamma$, $\rho \to \eta'+\gamma$ and others. Previous studies have placed a good constraint on the strange and light quark mixing angle $\phi$, and the result is: $\phi=39.3^o\pm1.4^o$. The latest Lattice QCD gives the mixing angle between gluonium and $c\bar{c}$, which is $\phi=(4.8^{+0.6}_{-1.0})^o$. In this work, we seek to estimate the values of other parameters $\theta_{g,c},a_{g,c}$ via the $J/\psi \to \eta^{(')}+\gamma$, $J/\psi \to \eta_c+\gamma$
and $J/\psi \to G+\gamma$ processes. Then we make theoretical predictions for processes such as radiative decays of other heavy quarkonium and electron-positron annihilation processes.

\begin{figure}[th]
\begin{center}
\includegraphics[width=0.4\textwidth]{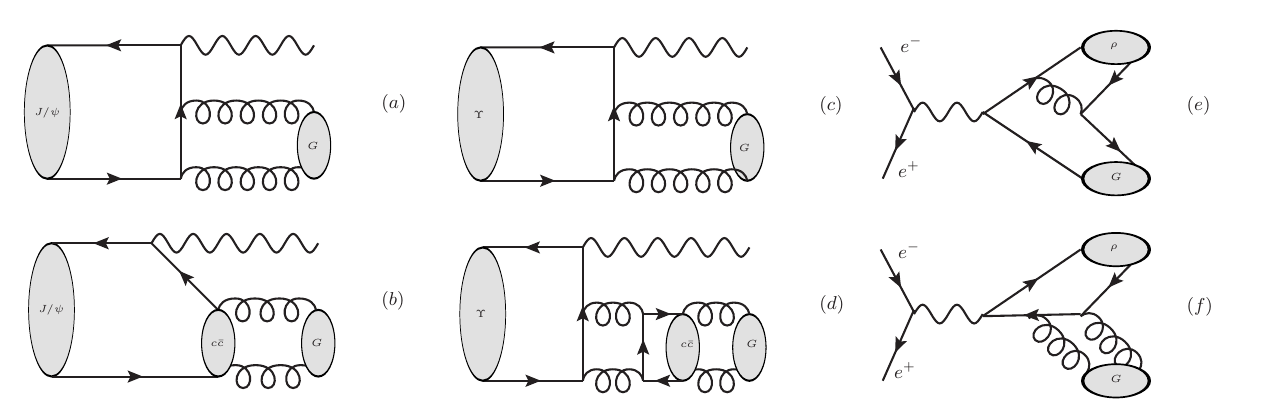}\,\,\,
\caption{Typical Feynman diagrams for $J/\psi$ and $\Upsilon$ radiative decays to a pseudoscalar Glueball, and electron-positron annihilation to a pseudoscalar Glueball and a light vector meson. Different topological diagrams appear in $J/\psi$ and $\Upsilon$ decays. }
\label{fig:feynmandiag}
\end{center}
\end{figure}

\textit{Pseudoscalar production in heavy quarkonium radiative decays and electron-positron annihilation.}
The typical Feynman diagrams for $J/\psi$ radiative decay to pseudoscalar Glueball is given in Fig.~\ref{fig:feynmandiag},
and other pseudoscalar meson involved process can be plotted similarly.
The Feynman amplitude of $J/\psi$ radiative decay to pseudoscalar meson at leading-power can be factorized into
\bqa
iM&=&eQ_c \langle 0|\chi^\dagger{\mbox{\boldmath $\sigma$}}\psi|J/\psi\rangle\int_0^1 du \int_0^1 dt \epsilon^{\alpha\beta\mu\nu}k_\alpha p_\beta\varepsilon_\mu\varepsilon^{'*}_\nu
 \nnb&&\times \mbox{\boldmath$H$}(u,t,\mu)
\mbox{\boldmath$\Phi$}(t,\mu)\,,\label{eq:amp}
\eqa
where $\mbox{\boldmath$H$}$ is the hard kernel
with two components. $Q_c$ is the charm quark charge. $\varepsilon(p)$ and $\varepsilon'(p-k)$ are the polarization
vectors of $J/\psi$ and the radiated photon, respectively.

The branching ratio of $J/\psi$ radiative decay to pseudoscalar meson can be written as
\begin{eqnarray}
{\cal B}(J/\psi\to P+\gamma)&=&\frac{m_{J/\psi}^2-m_P^2}{16\pi\Gamma_{J/\psi}
m_{J/\psi}^3}|\overline{M}|^2.\label{eq:Br}
\end{eqnarray}

Using this formulae, one can evaluate the remaining parameters in $4\times 4$ mixing scheme.
We list the inputting from PDG~\cite{ParticleDataGroup:2024cfk}: $m_{J/\psi}=3.0969$GeV, $\Gamma_{J/\psi}=92.9$keV, $m_{\eta_c}=2.984$GeV,
$m_{\eta}=0.548$GeV, $m_{\eta'}=0.958$GeV, $m_{X(2370)}=2.376$GeV.  We have treated the pseudoscalar Glueball as the dominant constituent of the $X(2370)$ to extract
its decay constant and mixing angles.
The branching ratios of $J/\psi$ radiative decay to pseudoscalar mesons are
\begin{eqnarray}
{\cal B}(J/\psi\to \eta_c+\gamma)&=&(1.82\pm0.15)\times 10^{-2},\nonumber\\
{\cal B}(J/\psi\to G+\gamma)&=&(2.3\pm0.8)\times 10^{-4},\nonumber\\
{\cal B}(J/\psi\to \eta'+\gamma)&=&(5.28\pm0.06)\times 10^{-3},\nonumber\\
{\cal B}(J/\psi\to \eta+\gamma)&=&(1.090\pm0.013)\times 10^{-3},
\end{eqnarray}
where the branching ratio for pseudoscalar Glueball is adopted from Lattice QCD~\cite{Gui:2019dtm}
and other data are from PDG~\cite{ParticleDataGroup:2024cfk}.

From the above experimental data, one can expect that the light quark-antiquark configuration can not explain
both $\eta$ and $\eta'$ data, whose amplitude is also suppressed in Fig.~\ref{fig:feynmandiag}. Thus
the $\eta-\eta'-\eta_c$ mixing is employed in many previous studies. In this work, we consider both the
$\eta-\eta'-\eta_c$  and $\eta-\eta'-G$ mixing.
One might expect these mixing effects to be small, since the mixing angles are relatively small.
However, due to their different physical masses, the phase space in $J/\psi$ decays differs,
and the kinematic factors may give rise to an enhancement effect.

The kinematic enhancement factor is
\begin{align}
K(m_i,m_j,m_n)=\left(\frac{k_j}{k_n}\right)^3=\left(\frac{m_i^2-m_j^2}{m_i^2-m_n^2}\right)^3,
\end{align}
where $m_i$ is the initial  $J/\psi$  mass; $m_{j,n}, k_{j,n}$ are the final pseudoscalar mass and momentum.
This kinematic enhancement factor is from both the phase space factor in Eq.~(\ref{eq:Br}) and the
antisymmetric tensor structure in the amplitude of Eq.~(\ref{eq:amp}). Take an example, the kinematic enhancement $K$ factors
are 2477 and 2015 for $\eta-\eta'-\eta_c$ mixing, which lead to large branching ratios for $\eta^{(')}$.
Though the kinematic enhancement $K$ factors
are 13 and 11 for $\eta-\eta'-G$ mixing, the larger mixing angle compared to that in $\eta-\eta'-\eta_c$ mixing will affect the final  branching ratios of $\eta^{(')}$.

\begin{figure}[th]
\begin{center}
\includegraphics[width=0.4\textwidth]{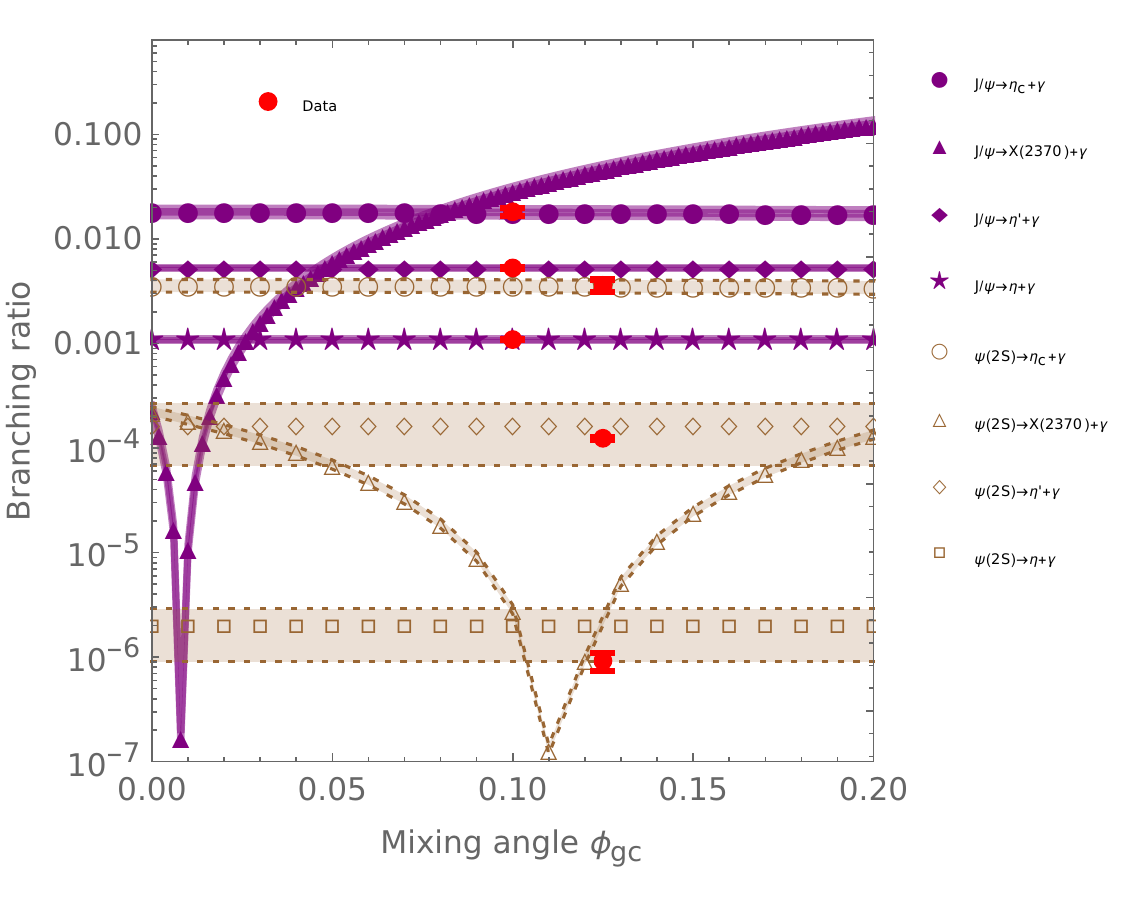}\,\,\,
\caption{Branching ratios for $J/\psi$ and $\psi(2S)$ radiative decays to a pseudoscalar meson.
The data are from PDG~\cite{ParticleDataGroup:2024cfk}.}
\label{fig:Brphi}
\end{center}
\end{figure}

Based on the $\eta-\eta'-G-\eta_c$ mixing scheme, $J/\psi$ radiative decay to pseudoscalar meson data and kinematic enhancement factors, the best-fit result for the remaining mixing angles is
\begin{align}
&I:~~a_c=0.0176,~\theta_c=-0.32;~a_g=-0.58,~\theta_g=-0.16.  \nonumber\\
&II:~a_c=-0.0176,~\theta_c=2.8;~a_g=0.58,~\theta_g=2.98.
\end{align}
There are two kinds of parameter values but give the same final results.

As a check of the previously proposed mixing scheme, we calculate the pseudoscalar meson production in $\psi(2S)$ radiative decays and compare them with experimental data. We plot our results for the branching ratios of $\psi(2S)$ radiative decays to a pseudoscalar meson in Fig.~\ref{fig:Brphi},
where our theoretical predictions are well agreement with experimental data.
The branching fraction for $X(2370)$ depends strongly on the mixing angle and exhibits significant variation. It is evident that destructive interference exists in both the $J/\psi$ and $\psi(2S)$ processes, however, the dependence of this destructive interference on the mixing angle differs between the two channels. Therefore, if the  $\psi(2S)$ radiative decays into $X(2370)$  is observed experimentally, it could determine the mixing angle and simultaneously provide a definitive result regarding the content of $X(2370)$.

In addition, we also apply our schemes to the $\Upsilon(nS)$ radiative decay processes, which have not yet been unambiguously measured experimentally.  For $\Upsilon$ radiative decays into the pseudoscalar meson, there are different topological diagrams from Fig.~\ref{fig:feynmandiag}. Thus measuring the branching fractions of these decay channels experimentally will undoubtedly provide a better test of the mixing scheme.  Here we give the final results as follows:
\begin{eqnarray}
{\cal B}(\Upsilon(1S,2S)\to \eta_c+\gamma)&=&(3\pm2,5\pm3)\times 10^{-5},\nonumber\\
{\cal B}(\Upsilon(1S,2S)\to G+\gamma)&=&(1.7\pm1.0,1.9\pm0.6)\times 10^{-6},\nonumber\\
{\cal B}(\Upsilon(1S,2S)\to \eta'+\gamma)&=&(5\pm3,5.2\pm3)\times 10^{-7},\nonumber\\
{\cal B}(\Upsilon(1S,2S)\to \eta+\gamma)&=&(9.1\pm5.4,8.7\pm5.2)\times 10^{-9}.\nonumber\\
\end{eqnarray}

\begin{figure}[th]
\begin{center}
\includegraphics[width=0.4\textwidth]{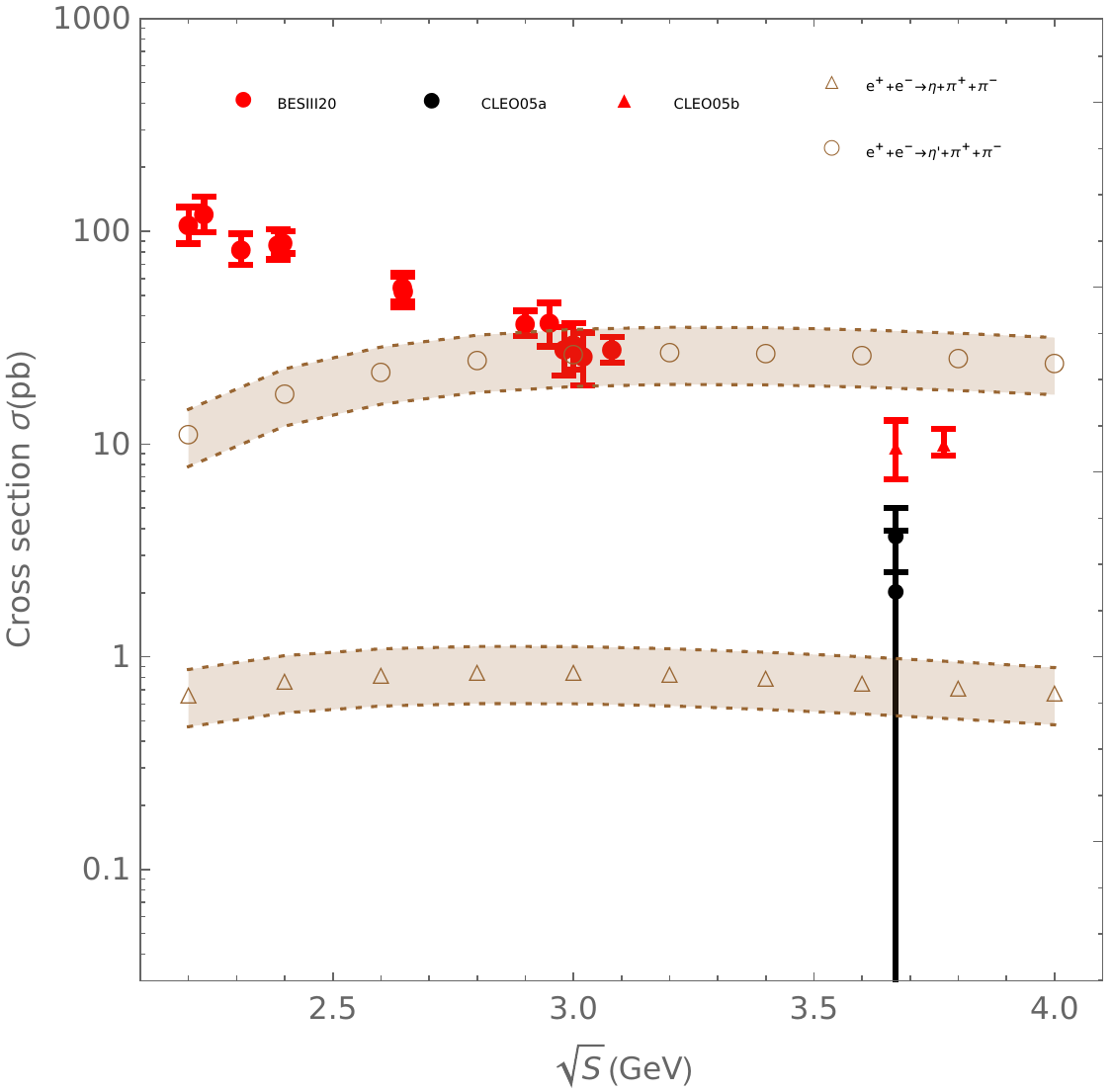}\,\,\,
\caption{Cross section(pb) for $e^+ e^- \to \eta^{(')}+\rho^0(\to \pi^++\pi^-)$. The BESIII data for $e^+ e^- \to \eta'+ \pi^++\pi^-$ process are from Ref.~\cite{BESIII:2020kpr}. The CLEO data for  $e^+ e^- \to \eta^{(')}+ \pi^++\pi^-$ process are from Ref.~\cite{CLEO:2005zrs}. }
\label{fig:crosssection}
\end{center}
\end{figure}

In the following, we will study the pseudoscalar production in electron-positron annihilation processes.
Their cross section can be written as
\begin{eqnarray}\label{cross-section}
	\sigma(e^+ e^- \to P+V) = \frac{4\pi \alpha^2}{3} \left( \frac{k_{P}}{\sqrt{S}} \right)^3 |F(S)|^2,
\end{eqnarray}
where $k_{P}$ is the magnitude of the three-momentum of the in the center-of-mass frame
\begin{eqnarray}
k_{P}=\frac{\lambda^{1/2}(S,m_{V}^2,m^2_P)}{2\sqrt{S}},
\end{eqnarray}
with the K\"allen function $\lambda(x,y,z)=x^2+y^2+z^2-2xy-2yz-2xz$.

Given that extensive long-term experimental studies on the $e^+ e^- \to \eta^{(')}+\rho^0(\to \pi^++\pi^-)$ process have already been conducted, we therefore perform a theoretical calculation of these processes. We give the final results in Fig.~\ref{fig:crosssection},
where our theoretical predictions for $e^+ e^- \to \eta'+\rho^0(\to \pi^++\pi^-)$  are well agreement with experimental data around 3GeV,
however, large deviates at both low and high energies, especially near 2GeV. We consider that hadronic resonances might play a crucial role, however, a systematic investigation of this issue is deferred to future studies.

For the $e^+ e^- \to X(2370)+\rho^0(\to \pi^++\pi^-)$ process, higher energies are required. We therefore investigated the production cross sections at several energy points
\begin{align}
&\sigma\left(\sqrt{S}=(4, 5, 10.58)GeV\right)\nonumber\\
=&\left(5.8\pm3.6, 11.7\pm7.0, 4.1\pm2.4\right)pb.
\end{align}
These results shall be tested in BESIII and Belle-II experiments.

\textit{Conclusion.}
In this work, we have treated the pseudoscalar Glueball as the dominant constituent of the $X(2370)$ and study the $J/\psi$ radiative decay to pseudoscalar meson in QCD factorization theory.  We have extracted the mixing angles in the $\eta-\eta'-G-\eta_c$  tetra-mixing scheme according to
the latest data for the $J/\psi$ radiative decay to pseudoscalar meson.  We then predict the decay rates for $\psi(2S)$ radiative decay to pseudoscalar meson and well-explain the experimental measurements, which in verse support the $X(2370)$ has a large pseudoscalar Glueball component. In addition, the $\Upsilon(1S)$ and $\Upsilon(2S)$ radiative decay  rates to pseudoscalar meson and the production cross section for pseudoscalar meson in electron-positron annihilation process are predicted, which  can be tested by current BESIII and Bell-II experiments.

\textit{Acknowledgments.}
The author thank the useful discussion with Prof. Cong-Feng Qiao and Prof. Wenlong Sang.
This work is supported by
the National Natural Science Foundation of China
Grants Nos.12322503 and 12235018.

\end{document}